\documentclass[twocolumn]{aastex61}

\usepackage{graphicx}
\usepackage{lineno}                   
\usepackage{amsmath}                  
\usepackage{dcolumn}                  
\usepackage{bm}                       
\usepackage{xspace}                   
\usepackage{hyperref}                 
\usepackage{gensymb}

\hypersetup{
  colorlinks=true,
  linkcolor=blue,
  citecolor=blue,
  urlcolor=blue,
}

\shorttitle{Search for Very High Energy Gamma Rays from the Northern \textit{F\small{ERMI}} Bubble Region}
\shortauthors{HAWC Collaboration}

\begin{document}
%
\title{Search for Very High Energy Gamma Rays from the Northern \textit{Fermi} Bubble Region with HAWC}

\correspondingauthor{H.A.~Ayala Solares}
\email{hayalaso@mtu.edu}

\author{A.U.~Abeysekara}
\affil{Department of Physics an Astronomy, University of Utah, Salt Lake City, UT, USA}

\author{A.~Albert}
\affil{Physics Division, Los Alamos National Laboratory, Los Alamos, NM, USA}

\author{R.~Alfaro}
\affil{Instituto de F\'{i}sica, Universidad Nacional Aut\'onoma de M\'exico, Mexico City, Mexico}

\author{C.~Alvarez}
\affil{Universidad Aut\'{o}noma de Chiapas, Tuxtla Guti\'{e}rrez, Chiapas, Mexico}

\author{J.D.~\'{A}lvarez}
\affil{Universidad Michoacana de San Nicol\'{a}s de Hidalgo, Morelia, Mexico}

\author{R.~Arceo}
\affil{Universidad Aut\'{o}noma de Chiapas, Tuxtla Guti\'{e}rrez, Chiapas, Mexico}

\author{J.C.~Arteaga-Vel\'{a}zquez}
\affil{Universidad Michoacana de San Nicol\'{a}s de Hidalgo, Morelia, Mexico}

\author{H.A.~Ayala Solares}
\affil{Department of Physics, Michigan Technological University, Houghton, MI, USA}

\author{A.S.~Barber}
\affil{Department of Physics and Astronomy, University of Utah, Salt Lake City, UT, USA}

\author{N.~Bautista-Elivar}
\affil{Universidad Politecnica de Pachuca, Pachuca, Hidalgo, Mexico}

% ...

\begin{abstract}
We present a search for very high energy gamma-ray emission from the Northern \textit{Fermi} Bubble region using data collected with the High Altitude Water Cherenkov (HAWC) gamma-ray observatory. The size of the data set is 290 days.
No significant excess is observed in the Northern \textit{Fermi} Bubble region, hence upper limits above 1\,TeV are calculated. 
The upper limits are between 3$\times 10^{-7}$\,$\text{GeV}\, \text{cm}^{-2}\, \text{s}^{-1}\, \text{sr}^{-1}$ and 4$\times 10^{-8}$\,$\text{GeV}  \,\text{cm}^{-2}\, \text{s}^{-1}\, \text{sr}^{-1}$.
The upper limits disfavor a proton injection spectrum that extends beyond 100\,TeV without being suppressed. 
They also disfavor a hadronic injection spectrum derived from neutrino measurements.
\end{abstract}

\section{Introduction}
\label{sec:Introduction}

The search for a counterpart of the microwave haze ~\citep{Dobler2010} in gamma-ray data, using the \textit{Fermi} Large Area Telescope (LAT), revealed the existence of two large structures extending up to $55\degree$ above and below the Galactic Plane~\citep{Dobler2010, Su2010}. Due to their bubble-like shape they received the name of \textit{Fermi} Bubbles. 

The gamma-ray emission of the \textit{Fermi} Bubbles presents a hard spectrum ---$dN/dE \sim E^{-2}$--- in the energy range from approximately 1\,GeV to 100\,GeV.
The surface brightness is roughly uniform in both bubbles ---except for a structure inside the South Bubble called  the cocoon--- and the total luminosity of the bubbles for galactic longitude $|b|>10\degree$ and between 100\,MeV and 500\,GeV was found to be $4.4^{+2.4}_{-0.9} \times 10^{37}$\,erg \,s$^{-1}$ ~\citep{Ackermann2014}.

The origin of the \textit{Fermi} Bubbles is still uncertain. Different models have been proposed to explain their formation. Most of the models revolve around the idea of outflows from the galactic center which then interact with the interstellar medium. The outflow can be generated by activity of the nucleus in our galaxy producing a jet~\citep{Guo2012a,Guo2012b}, wind from long time-scale star formation~\citep{Crocker2011}, periodic star capture processes by the supermassive black hole in the Galactic Center~\citep{Cheng2011}, or by winds produced by the hot accretion flow in Sgr A$^{\ast}$ ~\citep{Mou2015}.

The production of gamma rays is also under dispute.  Hadronic and leptonic models are the main mechanisms to explain the gamma-ray production. Photons of hadronic origin are due to the decay of neutral pions that are produced in the interaction of protons with nuclei in the interstellar medium (ISM). These protons are injected in the bubble regions by the outflow processes mentioned before or they can be accelerated inside the bubble as proposed by ~\cite{fujita2013,fujita2014}. Some of these models, ~\citep{Crocker2011,fujita2013}, predict the possibility of high-energy gamma rays. 
In the leptonic model, high-energy photons are produced by inverse Compton scattering from the interaction of energetic electrons with  photons from the interstellar radiation fields (IRF) or cosmic microwave background (CMB).  
The division between hadronic and leptonic models should not be strict, but rather, a combination of both models can be possible as shown in ~\cite{Cheng2011, Ackermann2014}.
Observations at other wavelengths, specifically at lower energies, have helped to constrain some models. For instance, the microwave haze ---produced by synchrotron radiation--- can help to constrain the electron population, which can also radiate in gamma rays ~\citep{Dobler2010,Su2010,Crocker2011,Mou2015,Guo2012b}.

The same principle can apply at very high energies (VHE; $>$100\,GeV), where observations can constrain the population of the highest-energy cosmic rays. 
Considering that the Northern \textit{Fermi} Bubble region is in the field of view of the High Altitude Water Cherenkov (HAWC), a search for VHE gamma rays (above 1\,TeV) is presented. The paper is divided as follows: The HAWC observatory and the data set used in the analysis are defined in Section \ref{sec:Detector}, the analysis procedure is described in Section \ref{sec:analysis}, and the results are discussed in Section \ref{sec:results} and summarized in Section \ref{sec:Conclusion}.

\section{The HAWC Observatory \& the Data Set}
\label{sec:Detector}
HAWC is a ground-based gamma-ray observatory. It is located between Volc\'{a}n Sierra Negra and Pico de Orizaba near Puebla, Mexico, at an altitude of 4100\,m a.s.l. and at $(18\degree 59'41"$N, $97\degree 18' 30"$W).
The observatory has a duty cycle of $>$95\% and a large field of view of $\sim$2\,sr, which allows it to cover $8.4$\,sr in a day~\citep{Abeysekara2013}. 
The instrument consists of an array of 300 water Cherenkov detectors (WCDs). The WCDs are steel tanks of 7.3\,m in diameter and 5\,m in height, filled with water up to 4.5\,m. 
Each WCD is filled with $\sim$ 200,000\, L of purified water. The array provides an effective area of $\sim$22,000\,m$^2$.
Inside the WCDs, four photomultiplier tubes (PMTs) facing upward are attached to the bottom. 

A simple multiplicity trigger is applied to find extensive air showers in the data. 
For the present analysis, the trigger requires 28 PMTs detecting Cherenkov light within a 150\,ns time window to be activated in order to identify a shower event.
After the processing and calibration of the events, the air shower cores, footprint brightness in the array, and gamma- and cosmic-ray directions are reconstructed. 
More information on the detector, calibration, and reconstruction is presented in ~\cite{CrabPaper}.

The HAWC observatory began science operations in August 2013, when it was still under construction. The analysis described in this paper uses data between 2014 November 27th to 2016 February 11th.  

The data set is divided into seven event-size bins represented by the fraction \textit{f} of functioning PMT channels triggered in an air shower event. The energy of the observed gamma rays is related to the shower event size that is measured in the HAWC array. The range of \textit{f} for this analysis goes from 0.162 to 1.00. Table \ref{table:analysisbins} shows the ranges for each analysis bin.

\begin{deluxetable}{cc}
\tablewidth{6cm}
\tablecaption{Analysis Bins defined as the fraction of operational PMT channels triggered in an air shower event. \label{table:analysisbins}}
\tablehead{
\colhead{ Analysis Bin } & \colhead{ $f$}}
\startdata
  $f_1$  & 0.162 - 0.247  \\ 
  $f_2$ & 0.247 - 0.356  \\
  $f_3$ & 0.356 - 0.485 \\
  $f_4$ & 0.485 - 0.618 \\
  $f_5$ & 0.618 - 0.740 \\
  $f_6$ & 0.740 - 0.840\\
  $f_7$ & 0.840 - 1.00 \\
\enddata
\end{deluxetable}

Standard selection cuts are applied to the data that pass the trigger condition. The signals in each PMT are required to have $>$1 photoelectrons (PEs) and are required to be between 150\,ns before and 400\,ns after the trigger.
In addition, it is required that more than 90\% of the PMT channels are functioning during the observation. Finally, cuts are applied to distinguish between gamma rays  and hadronic cosmic rays, the latter being the main background of measurements with the HAWC observatory. 
All the cuts are optimized by studying the Crab Nebula in the HAWC data ~\citep{CrabPaper}.

\section{Analysis}
\label{sec:analysis}

The analysis is focused on measuring the flux of gamma rays in the Northern Fermi Bubble Region, since this region is inside the field of view of HAWC. The main challenge of the analysis is to estimate the background of the data set. The procedure to deal with the background is divided in three steps:
\begin{enumerate}
\item Distinguish the air shower signatures between cosmic rays and gamma rays. The gamma-hadron cuts select the gamma-like showers.
\item Find the isotropic flux of cosmic rays and gamma rays. This is found by using direct integration and it is explained in Section \ref{subsec:map}.
\item Remove effects of the large-scale anisotropy as seen in \citep{Abeysekara2014}. The procedure is shown in Section \ref{subsec:excess}.
\end{enumerate}

\subsection{Direct Integration Isotropic Background Estimation}
\label{subsec:map}
The positions of the events are binned in equatorial coordinates using the HEALPix scheme~\citep{Gorski2005}. These are referred to as sky maps. For the analysis we set the pixel size to be $\sim 0.11\degree$.

The isotropic background is estimated using the direct integration (DI) technique described in~\cite{Atkins2003}. The background is integrated over 24 hours and therefore only data were used when the detector performance was stable for 24 hours, since this is a requirement for the integration technique. This results in a lifetime for the analysis of 290 days. 

As explained in ~\cite{Abeysekara2014}, an analysis based on a background integration period of $\delta t$ is sensitive to potential signal excesses of an RA size smaller than   $\delta t \, \cdot 15\degree$\,hour$^{-1}$.  Using a 24 hour integration period ensures that the analysis is sensitive to the Fermi Bubbles which extend to $\sim50\degree$ in RA. 

\begin{figure}[!h]
  	\centering
    \includegraphics[width=0.45\textwidth]{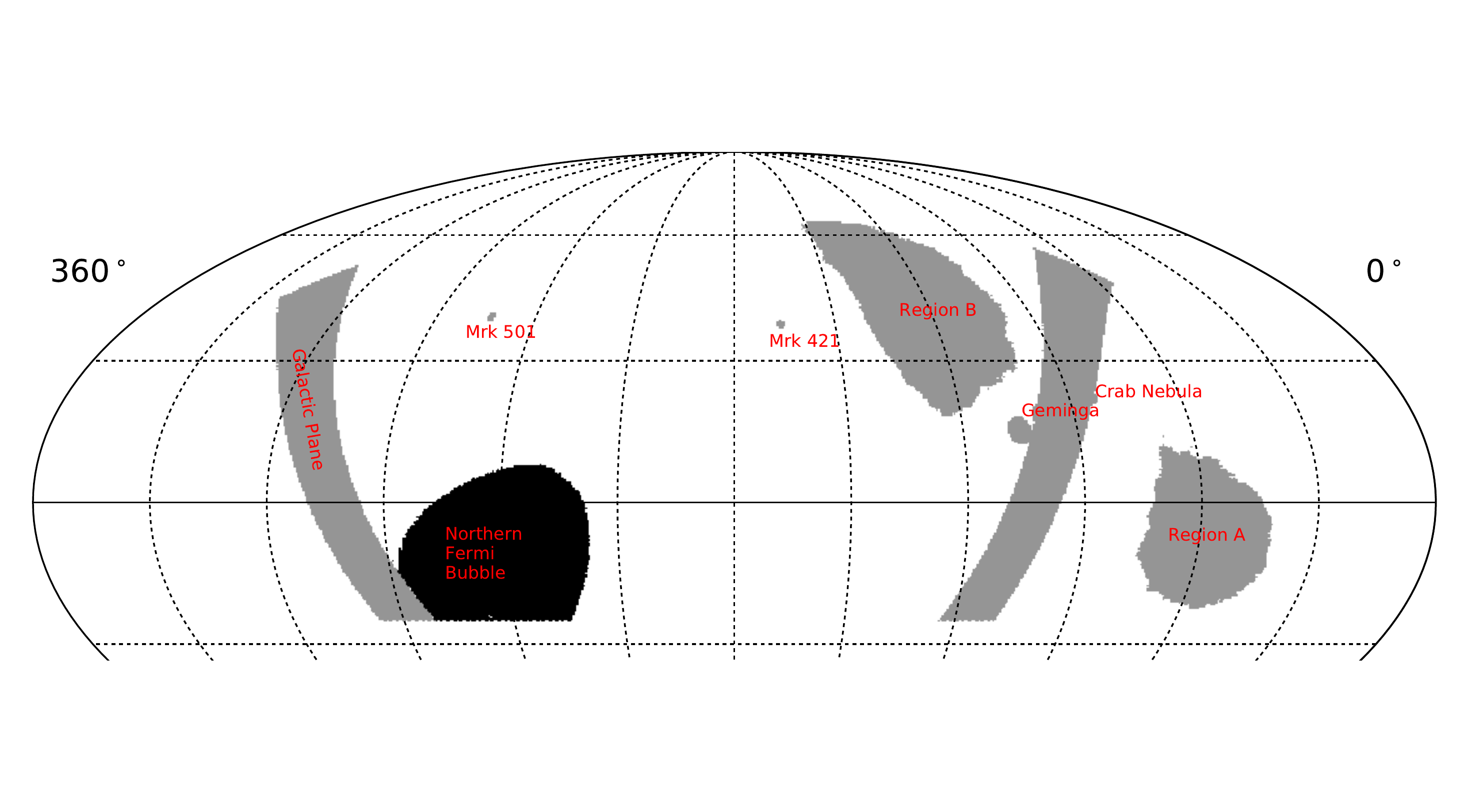}
    \caption{Skymap in equatorial coordinates showing the region of interest masking as used for the analysis. The masked ROI include the Galactic Plane, Geminga, the Crab Nebula, Mrk 421, Mrk 501, the small-scale anisotropies region A and B, and the Northern \textit{Fermi} Bubble.}
    \label{fig:mask}
\end{figure}

Since the estimation of the isotropic background can be biased by strong known sources in the data, a \textit{region of interest} (ROI) masking is used, as shown in Figure \ref{fig:mask}. The ROI masking covers the galactic plane [$\pm 6\degree$], as well as circular regions of radius $3\degree$, $1.3\degree$,$1\degree$ and $1\degree$, respectively, for Geminga, the Crab Nebula, Mrk 421 and 501. Region A and B from the small-scale cosmic-ray anisotropy are also masked. Their shapes are obtained from the results in~\cite{Abeysekara2014}, by requiring that the significances in the sky map without gamma-hadron cuts are greater than 4$\sigma$. Finally, the ROI for the Northern \textit{Fermi} Bubble was obtained from the \textit{Fermi} Diffuse Model pass 7 version 6 \footnote{See \url{http://fermi.gsfc.nasa.gov/ssc/data/access/lat/BackgroundModels.html}}. 

The shape of the Northern \textit{Fermi} Bubble above 1\,TeV is unknown. We perform a gamma-ray flux excess search within the boundaries of the Northern Bubble as detected by \textit{Fermi} below TeV energies.

\subsection{Gamma-Ray Excess Calculation}
\label{subsec:excess}

For each analysis bin, sky maps are created after applying the gamma-hadron cuts. The isotropic background is then estimated and an excess sky map is obtained. The excess in each pixel $i$ is given by the following equation,
\begin{equation} \label{eq:exeff}
E'_i = N'_i - \langle N'_i \rangle,
\end{equation}
where $N'_i$ is the observed data after gamma-hadron cuts, and $\langle N'_i \rangle$ is the isotropic background estimated after gamma-hadron cuts.
However,  the excess sky map in the lower analysis bins reveals the large-scale anisotropy as seen in ~\citep{Abeysekara2014}. This is because there is enough statistics to calculate the background with an accuracy of one part per mille in these bins.
The Northern \textit{Fermi} Bubble is located at a deficit region, therefore this systematic effect needs to be removed. The subtraction of this cosmic-ray feature is achieved  by  using the data without gamma-hadron cuts.

The data without gamma-hadron separation is composed of a total number of gamma rays and cosmic rays,
\begin{equation} \label{eq:data}
N_i = G^T_i + C^T_i 
\end{equation}
where $N_i$ is the data without gamma-hadron cuts in the pixel $i$, $G^T_i$ is the number of gamma rays in the pixel $i$, and $C^T_i$ is the number of cosmic rays in the pixel $i$. $G^T_i$ and $C^T_i$ can also be decomposed in terms of an isotropic component and an excess(or deficit). This is expressed as
\begin{eqnarray}
G^T_i &= & G^I_i + G_i \nonumber \\
C^T_i &= & C^I_i + C_i ,
\end{eqnarray}
where $G^I_i$, $C^I_i$ are the isotropic components; and $G_i$, $C_i$ are the gamma-ray and cosmic-ray excesses or deficits.

The data after gamma-hadron separation also contains gamma rays and cosmic rays but the composition is different due to the rejection efficiency of the gamma-hadron separation cuts,
\begin{equation} \label{eq:dataeff}
N'_i = \varepsilon_{G,i} G^T_i + \varepsilon_{C,i}C^T_i,
\end{equation}
where $\varepsilon_{G,i}$ and $\varepsilon_{C,i}$ are the gamma and hadron efficiencies after applying the gamma-hadron cuts.

For completeness the isotropic background for the data before and after gamma-hadron cuts are written as follows,
\begin{eqnarray}
\langle N_i \rangle = G^I_i + C^I_i  \label{eq:bkg}  \\
\langle N'_i \rangle= \varepsilon_{G,i} G^I_i + \varepsilon_{C,i}C^I_i.\label{eq:bkgeff}
\end{eqnarray}

The gamma passing rate efficiency $\varepsilon_{G,i}$ is obtained using simulations. The detector response is simulated in each of the seven analysis bins and for 5$\degree$ declination bands between $-37.5\degree$ and $77.5\degree$. Each bin contains an energy histogram that is expected for the simulated signal. We compute the number of events in the energy histograms and the ratio for the events with gamma-hadron cuts $h'(e)$ over the events with no gamma-hadron cuts $h(e)$, where $e$ is the energy. Therefore, the efficiency can be written as
\begin{equation}
\varepsilon_{G,i} = \frac{\int h'(e)de}{\int h(e)de}.
\end{equation}

The hadron passing rate efficiency $\varepsilon_{C,i}$ is estimated from the data since the total number of cosmic rays outnumbers the total number of gamma rays. In order to avoid bright sources, we use equations \ref{eq:bkg} and \ref{eq:bkgeff} to estimate $\varepsilon_{C,i}$,
\begin{equation}
\varepsilon_{C,i} = \langle N'_i \rangle/\langle N_i \rangle. 
\end{equation}
The isotropic gamma-ray emission can be safely neglected in the sums for equations \ref{eq:bkg} and \ref{eq:bkgeff}. To account for declination dependence, the data in the same declination (or HEALPix ring) as in pixel $i$ is used so that $\varepsilon_{C,i} = \sum_j \langle N'_j \rangle / \sum_j \langle N_j \rangle$, where $j$ is the pixel in the ring containing $i$.

By combining Equations \ref{eq:dataeff} and \ref{eq:bkgeff}, Equation \ref{eq:exeff} can be re-written as
\begin{equation}
E'_{i} = \varepsilon_{C,i}C_i + \varepsilon_{G,i}G_i. \label{eq:effmet1}
\end{equation}
Using Equations \ref{eq:data} and \ref{eq:bkg} give an equation for the excess in pixel $i$  for the data without gamma-hadron cuts,
\begin{equation} \label{eq:effmet2}
E_{i} = C_i + G_i. 
\end{equation}


Finally, the number of gamma rays is obtained from Equations \ref{eq:effmet1} and \ref{eq:effmet2}, 
\begin{equation}
G_i=\frac{E'_i -\varepsilon_{C,i} E_i}{\varepsilon_{G,i}-\varepsilon_{C,i}}.
\end{equation}

The efficiency $\varepsilon_{G,i}$ is applied to the number of gamma rays $G_i$ to obtain the number of excess events measured by the detector.
\begin{equation}\label{eq:excess}
G'_i = \varepsilon_{G,i} G_i
\end{equation}

The previous equation is used to calculate the number of gamma rays G in each pixel inside the Northern Bubble region as defined in Figure \ref{fig:mask} and then summed to get a total excess $G' = \sum_i G'_i$ in each analysis bin. 

As mentioned in Section \ref{subsec:map}, the shape of the Fermi Bubbles at high energies is unknown, though some authors suggest that the size of the bubbles increases with energy \citep{fujita2013,Yang2014,Mou2015}.
In this case, calculating the flux in the smaller region of the MeV-GeV excess is the more conservative approach.

The description of the variables is presented in appendix \ref{app:variables}.

The uncertainty calculation for $G_i$ is shown in appendix \ref{app:uncertainty}.



\subsection{Testing the Analysis Method}
\label{subsec:simulation excess}
\begin{figure}[!ht]
  \centering
    \includegraphics[width=0.5\textwidth]{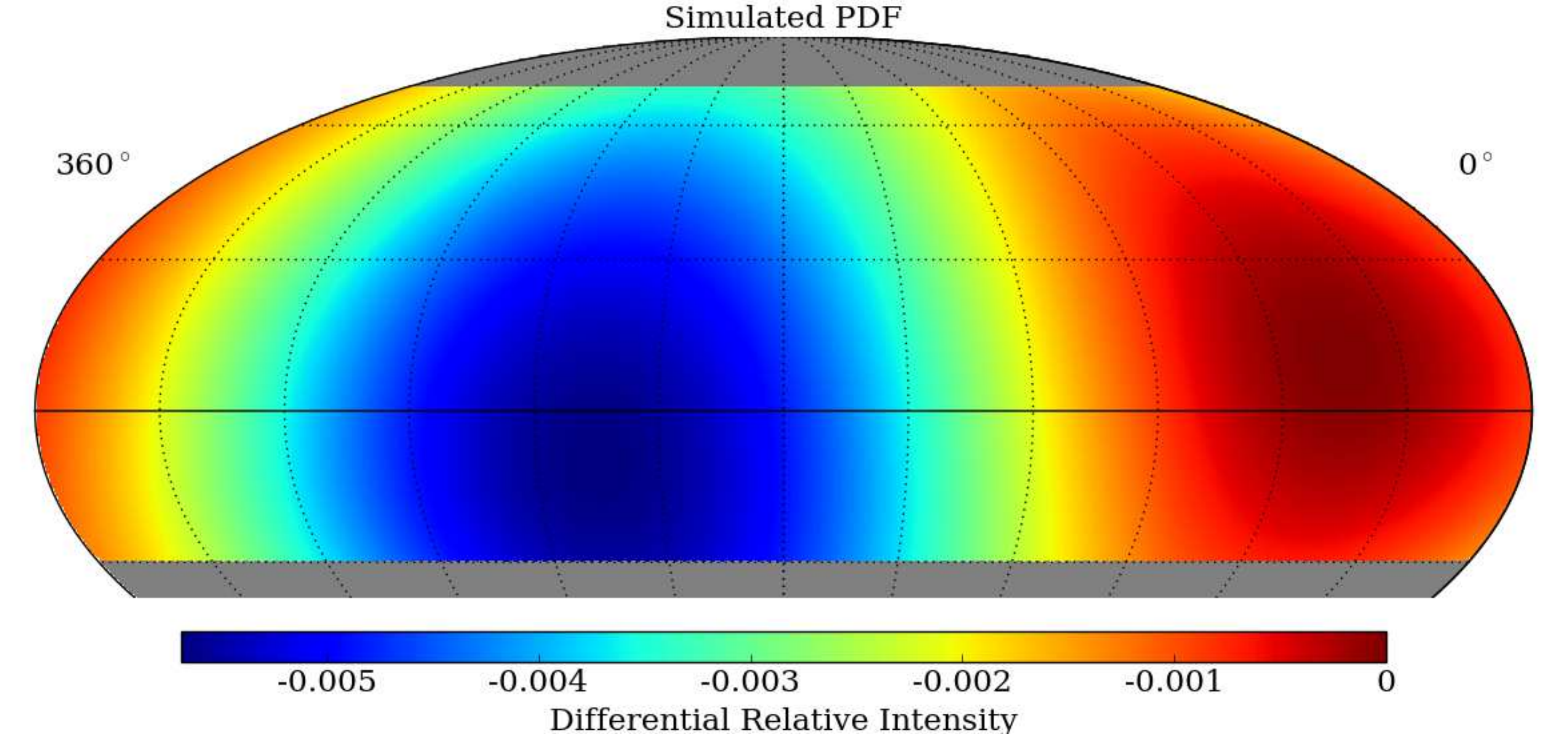}
    \caption{Dipole distribution used for the skymap simulation.}
    \label{fig:CRPDF}
\end{figure}

\begin{figure}[!ht]
	\centering
	\epsscale{0.8}
	\includegraphics[width=0.5\textwidth]{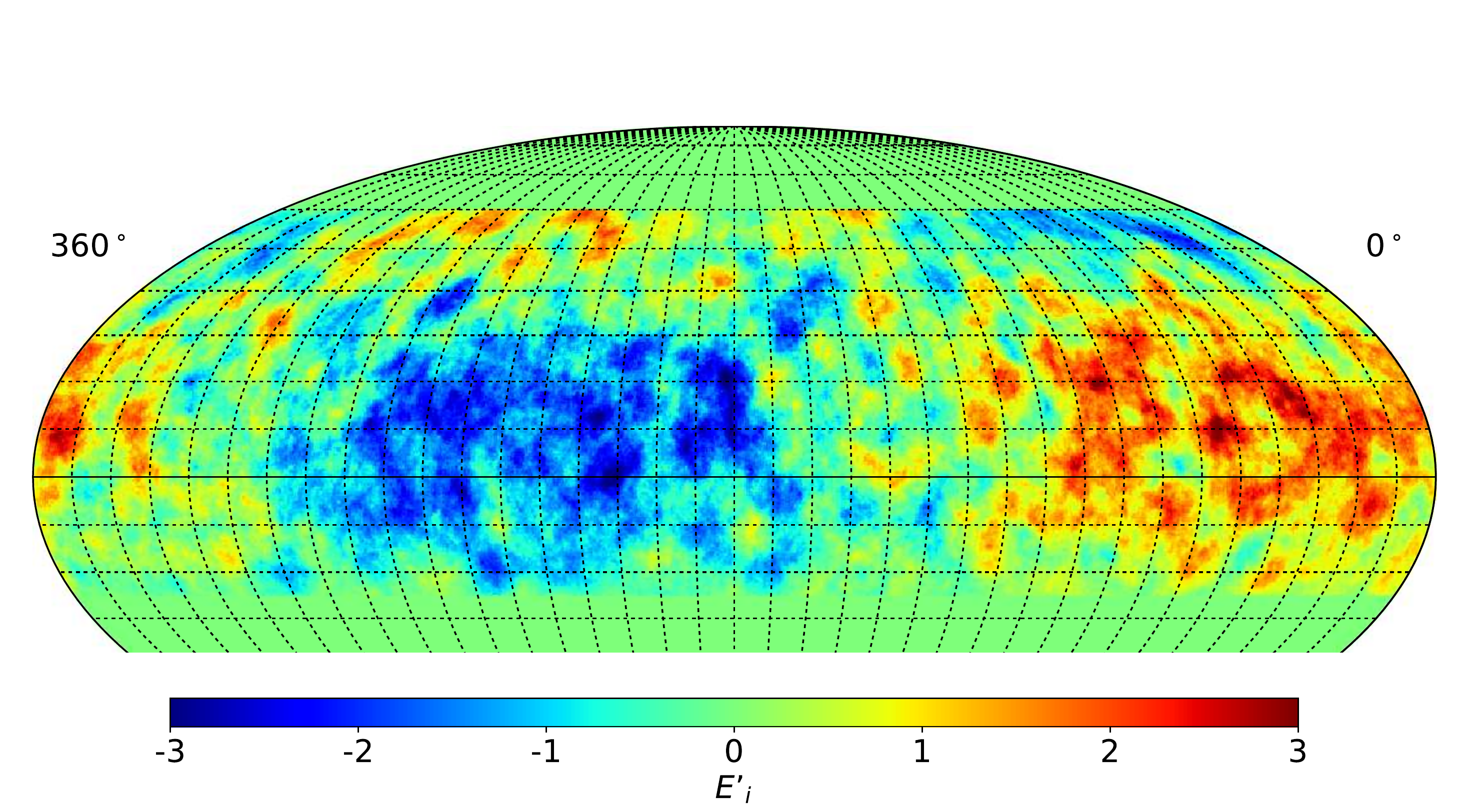}
	\includegraphics[width=0.5\textwidth]{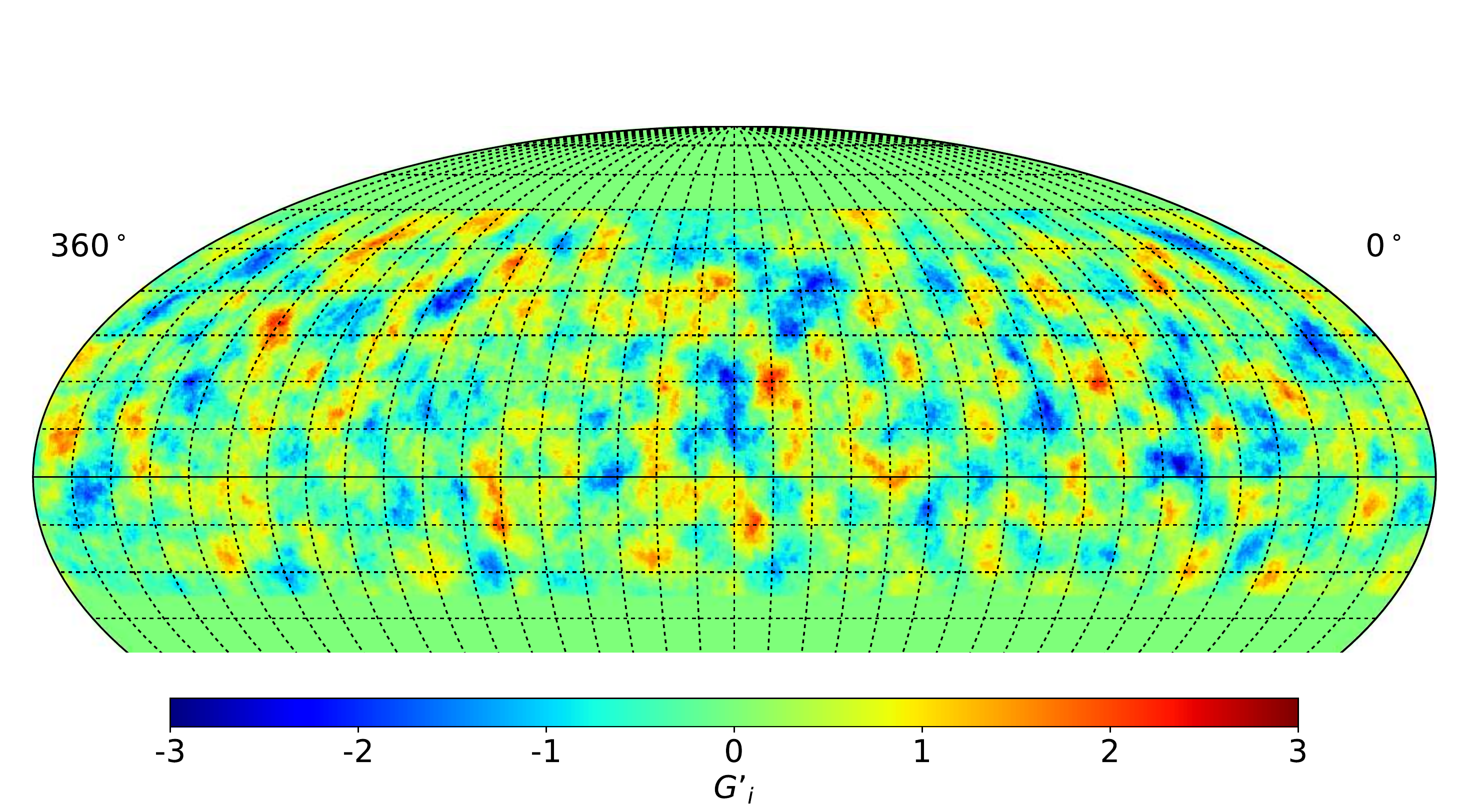}
	\caption{Simulated event excess in \textit{f}$_1$ for an injected signal which mimics the cosmic-ray anisotropy, smoothed with a 5\degree tophat. \textit{Top}: Event excess after subtracting the estimated background from the simulated data. \textit{Bottom}: Large-scale structure dominated by cosmic rays is removed after the method described in Section \ref{subsec:excess} is applied.}
	\label{fig:simexcessmap}
\end{figure}

The analysis method is tested on simulated sky maps containing a dipole distribution as shown in Figure \ref{fig:CRPDF} assuming no sources are present. 
A rate map in the local coordinates of HAWC containing a snapshot of 24\,sec of data is generated. Since HAWC observations cover a local sky of zenith angles $0\degree<\theta<45\degree$, the rate map is generated for this zenith angle range. Using the dipole distribution given in Figure \ref{fig:CRPDF}, the total sky event rate from HAWC data, and information from the detector response, a rate in each pixel is obtained. After the 24\,sec period the rate map is reset and the procedure is started again. In this way a simulated data set is generated that is of the same size as the real data set analyzed in this paper for both cases of without and with gamma-hadron cuts. 
An example of a resulting simulated sky map $f_1$ is shown in Figure \ref{fig:simexcessmap}. The upper panel presents the resulting map after simply subtracting the estimated background from the data, the lower panel shows the excess map after applying the procedure described in Section \ref{subsec:excess}.

\begin{figure}[!ht]
	\hspace*{-0.63cm} 
	\includegraphics[width=0.55\textwidth]{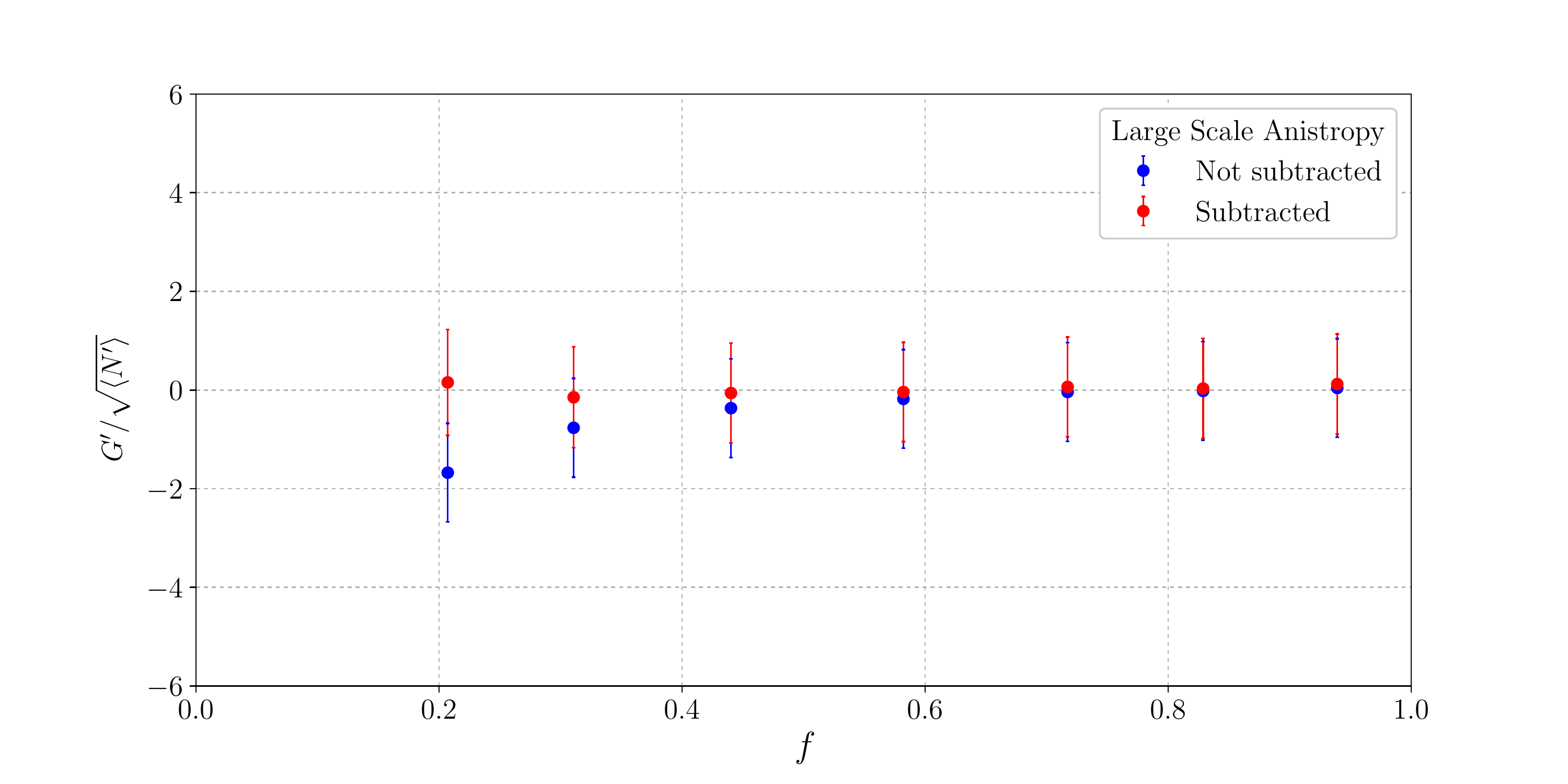}
	\caption{Simulated event excess over the square-root of the isotropic background inside the Northern Bubble region. The effect of the dipole is stronger at lower values of \textit{f}. Blue: Estimated DI background subtracted from simulated data; Red: Background subtraction considering  the gamma and cosmic-ray efficiencies. See method described in Section \ref{subsec:excess}.}
	\label{fig:simexcess}
\end{figure}

Figure \ref{fig:simexcess} shows the resulting simulated excesses in each $f$ bin. A comparison is made between the event excesses derived from simply subtracting the estimated DI background from the simulated data (blue points) and the event excess obtained by the method from Section \ref{subsec:excess} (red points). The effect of the simulated cosmic-ray anisotropy results in systematically lower excesses for the lowest two $f$ bins if the method from Section \ref{subsec:excess} is not applied.

The method is also tested by adding a strong \textit{Fermi} Bubble-like gamma-ray emission. The spectrum is assumed to be a power-law with spectral index $\gamma=2.0$ and normalization of $5.03\times10^{-7}$\,GeV$^{-1}$\,cm$^{-2}$\,s$^{-1}$\,sr$^{-1}$, both values obtained by fitting the \textit{Fermi} data points in the range of 1\,GeV to 150\,GeV. 

Using this assumption and extending the spectrum to TeV energies, the analysis procedure was tested. If the Northern \textit{Fermi} Bubbles had this spectrum, the HAWC observatory would have detected it.

\begin{figure}[!ht]
	\hspace*{-0.65cm} 
	\includegraphics[width=0.55\textwidth]{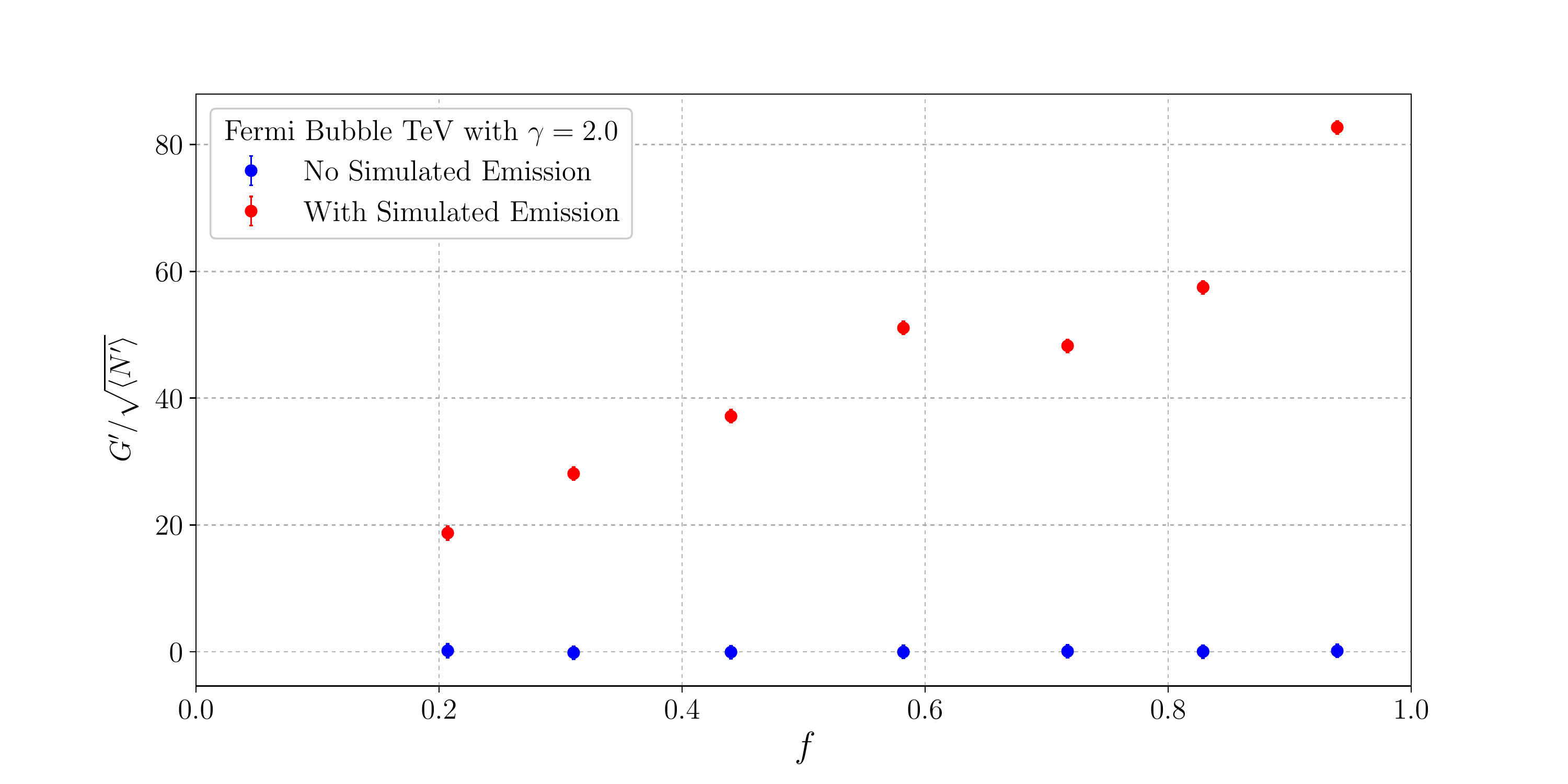}
	\caption{Simulated event excess over the square-root of the isotropic background inside the Northern Bubble region without and with strong \textit{Fermi} Bubble-like emission after applying the procedure. See method described in Section \ref{subsec:excess}.}
	\label{fig:simexcess2}
\end{figure}

\section{Results and Discussion}
\label{sec:results}

\subsection{Gamma-Ray Excess Results}
\begin{figure}[!ht]
	\centering
	\epsscale{0.8}
	\includegraphics[width=0.5\textwidth]{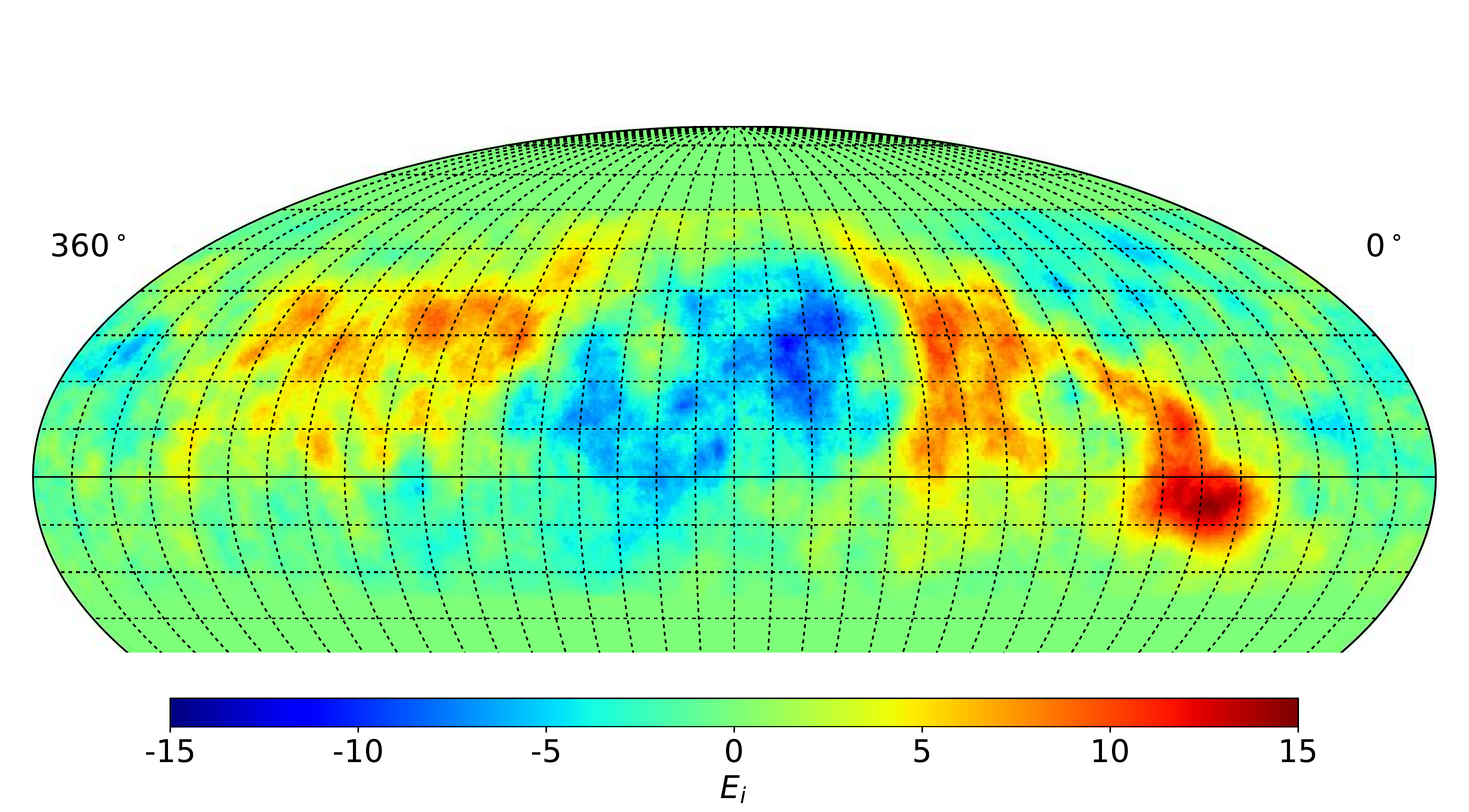}
	\includegraphics[width=0.5\textwidth]{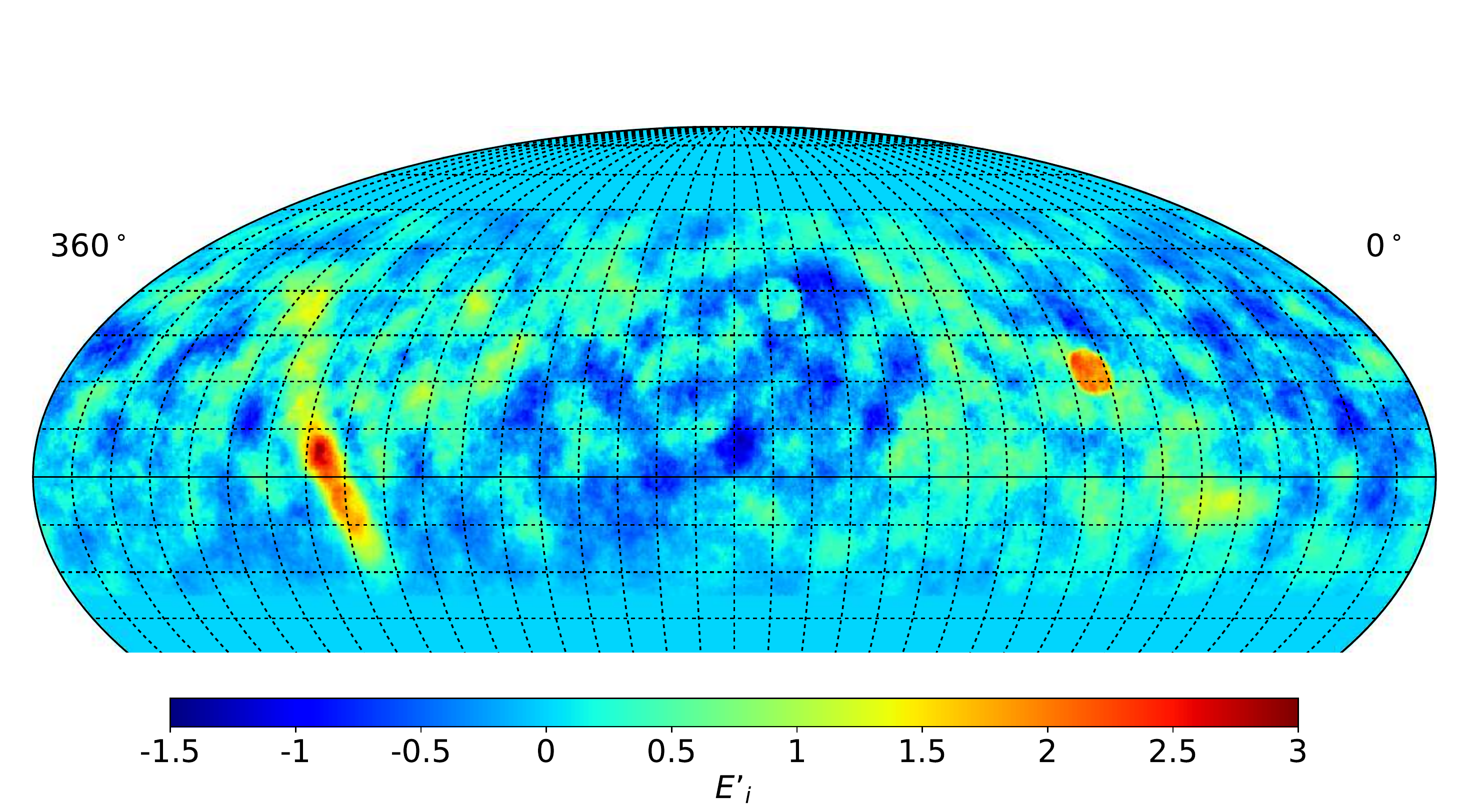}
	\includegraphics[width=0.5\textwidth]{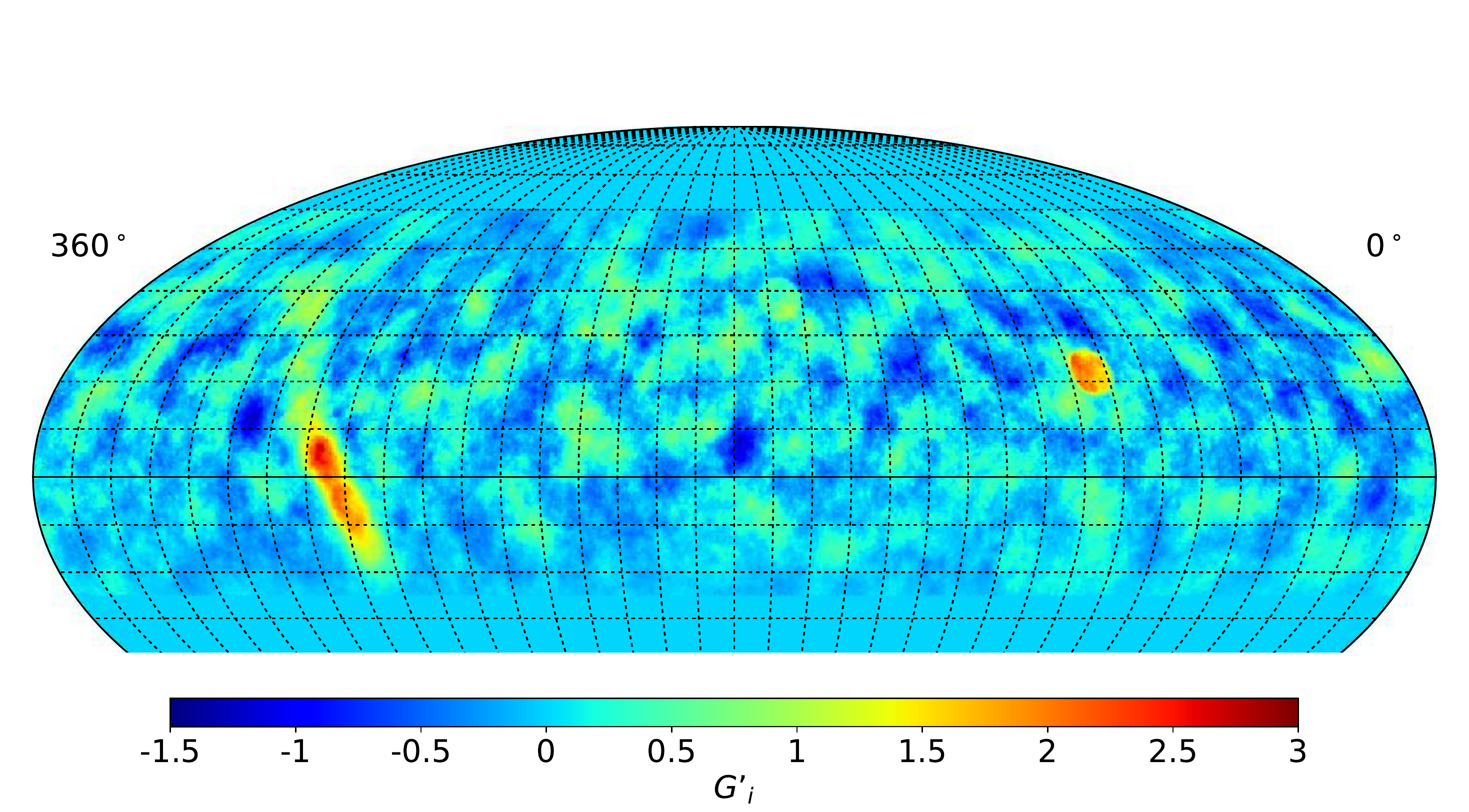}
	\caption{Event excesses in analysis bin \textit{f}$_1$, smoothed with a $5\degree$ tophat. \textit{Top}: Event excess $E_i$ after subtracting the estimated DI background from the cosmic-ray data. The large-scale CR anisotropy is visible. \textit{Middle}: Event excess $E'_i$ after subtracting the estimated DI background from the gamma-ray data. A deficit casued by the large-scale anisotropy is visible. \textit{Bottom}: Large-scale CR anisotropy structure is removed after the method described in Section \ref{subsec:excess} is applied.}
	\label{fig:excessmap}
\end{figure}

Figure \ref{fig:excessmap} shows skymaps of the result of the first analysis bin, \textit{f}$_1$. The figure shows a sky map without gamma-hadron cuts, and sky maps with gamma-hadron cuts before and after applying our procedure. 

\begin{figure}[!ht]
	\hspace*{-0.75cm} 
	\includegraphics[width=0.55\textwidth]{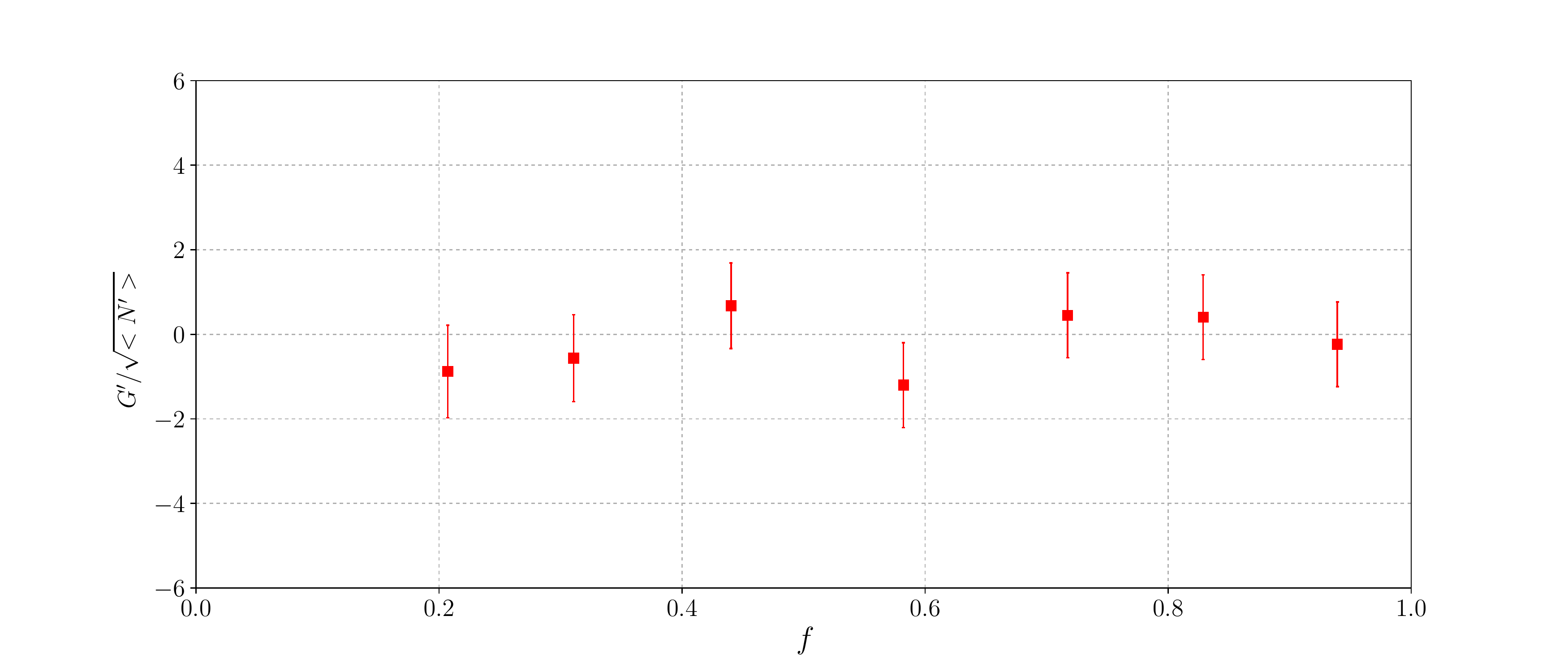}
	\caption{Event excess over the square-root of the isotropic background  inside the Northern Fermi Bubble region after applying the procedure described in Section \ref{subsec:excess}.}
	\label{fig:excess}
\end{figure}

Figure \ref{fig:excess} shows the results of the summed excess inside the bubble region after applying our procedure. No significant excess inside the bubble region is observed, therefore upper limits on the differential flux are calculated. The upper limits are also compared to the differential sensitivity of the HAWC observatory.
The upper limits give the maximum flux intensity that is plausible given the observed counts in the HAWC data. The differential sensitivity quantifies the power of the detection procedure and is based on  finding an $\alpha$-level threshold (related to background fluctuations claimed as detections) and the probability $\beta$ to detect a source\footnote{The definitions of upper limit and sensitivity are the same as upper bound and upper limit in \cite{Vinay2010}.}.

\subsubsection{Calculating the Upper Limits}\label{subsubsec:weights}
\begin{figure}[!ht]
	\hspace*{-0.63cm} 
	\includegraphics[width=0.55\textwidth]{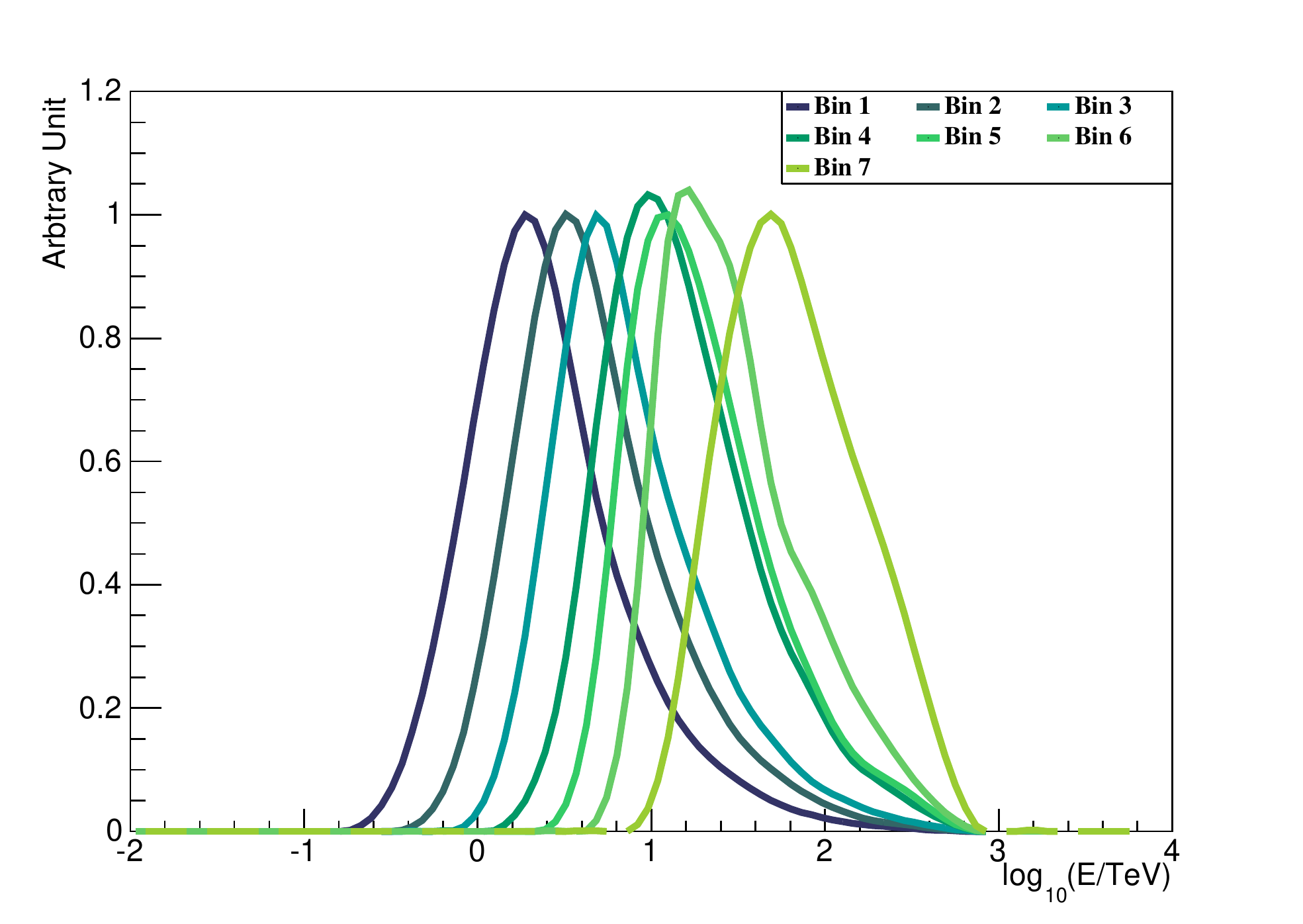}
	\caption{Energy distributions for the analysis bins assuming a 2.75 power-law spectrum.}
	\label{fig:energies}
\end{figure}

The differential flux is calculated from the measured excess by comparing the signal observed in the data  to an expected signal obtained for each of the fractional $f$ analysis bins using simulations. Since the energy response histograms for each analysis bin overlap (see Figure \ref{fig:energies}), the excesses measured in the analysis bins are combined in a weighted sum.

The procedure is as follows: a differential flux is assumed in an energy bin of width $\Delta \log(E/1\,\text{TeV})$. The width of differential energy bins is defined such that the results are independent of spectral assumptions. 
Using the HAWC detector response, an expected signal for the Northern Bubble region  is obtained for each fractional $f$ bin. Taking into account the previous values, the weight in the energy bin $k$ for the fractional bin $f_l$ is calculated as
\begin{equation}
w^k_l = \frac{ME^k_l}{\langle N' \rangle_l},
\end{equation}
where $w^k_l$ is the weight in the energy bin $k$ for the fractional $f_l$; $ME^k_l$ is the expected signal in the energy bin $k$ for the fractional $f_l$, and $\langle N' \rangle_l$ is the background estimated in fractional bin $f_l$. This procedure results in a matrix that allows ``projection" of the fractional $f$ analysis bin space onto the energy space.

Using the weights, the ratio of the observed signal and the expected signal is calculated
\begin{equation}
\label{eq:ratio}
R_k  =  \frac{\sum\limits^{f}_{l=1} w^k_l G'_l}{\sum\limits^{f}_{l=0} w^k_l ME^k_l}, 
\end{equation} 
and the uncertainty in the ratio as
\begin{equation}\label{eq:ratioerr}
\delta R_k  =  \frac{\sqrt{\sum\limits^{f }_{l=1} (w^k_l \delta G'_l)^2}}{\sum\limits^{f}_{l=1} w^k_l ME^k_l}.
\end{equation}
The ratio is used to obtain an estimation of the flux in the energy bin $k$
\begin{equation} \label{eq:flux}
F_k = (R \pm \delta R)_k F(E_k),
\end{equation}
where $F(E_k)$ is the flux assumption at the energy bin $k$ used to obtain the expected excess $ME^k_l$.

The upper limit calculation is then performed in the energy bins. 
The prescription of ~\cite{Helene1983} is used to calculate an upper limit on the differential flux derived from equation \ref{eq:flux}. A 95\% confidence level is chosen.

\subsubsection{Calculating the Sensitivity}
The sensitivity is calculated based on \cite{Vinay2010}\footnote{Named upper limit in the reference instead of sensitivity}. The procedure consists of setting a small probability for false positives (Type I error) and setting a probability of detection when there is a source (related to Type II error). The probability for false positives is set to  $\alpha=0.05$ and the probability of detection is set to  $\beta=0.5$.  This is to compare the detection power of the observatory to the calculated upper limit at 95\% confidence level. 

The calculation is performed by using the measured background and doing simulations for a simulated \textit{Fermi} Bubble of varying flux. For the simulation we assume a power-law with an index of -2.75 in the differential energy bin. For each analysis bin, the total background counts and the total expected number of events from the simulated source are calculated inside the bubble region. Following the same procedure as in Section \ref{subsubsec:weights}, the analysis bins are combined to get the total number of events for each energy bin. In each energy bin, a null hypothesis histogram and an alternative hypothesis histogram are created for the quantity 
\begin{equation}
S_k =  E_k /\sqrt{\left\langle N'\right\rangle_k}, 
\end{equation}
where $E_k$ is obtained by Poisson-fluctuating $\left\langle N' \right\rangle_k$ for the null hypothesis, and then subtracting $\left\langle N' \right\rangle_k$ from this value; or by Poisson-fluctuating $\left\langle N' \right\rangle_k+ME_k$  for the alternative hypothesis, and then subtracting $\left\langle N' \right\rangle_k$ from this value.   The Poisson fluctuations are performed 10000 times to fill the histograms. The null hypothesis histogram is used to find the $\alpha$-level detection threshold and the alternative hypothesis histogram is used to find the flux normalization that is required to obtain a probability of  detection of 0.5. 

\subsection{Differential Flux of the \textit{Fermi} Bubbles}
\label{subsec:upperlimit}
The first energy bin is centered at 2.2\,TeV, which is the median energy of fraction \textit{f}$_1$ assuming a power law spectrum of index $\gamma=2.75$ (see Figure \ref{fig:energies}). The energy bin width is set to $\Delta \log(E/1\,\text{TeV})=0.5$ which is comparable to the width of the energy histograms. 
The energy range covers up to the highest energy at which HAWC is sensitive ($\sim$100\,TeV).

\begin{figure*}
	\centering
	\includegraphics[width=0.95\textwidth]{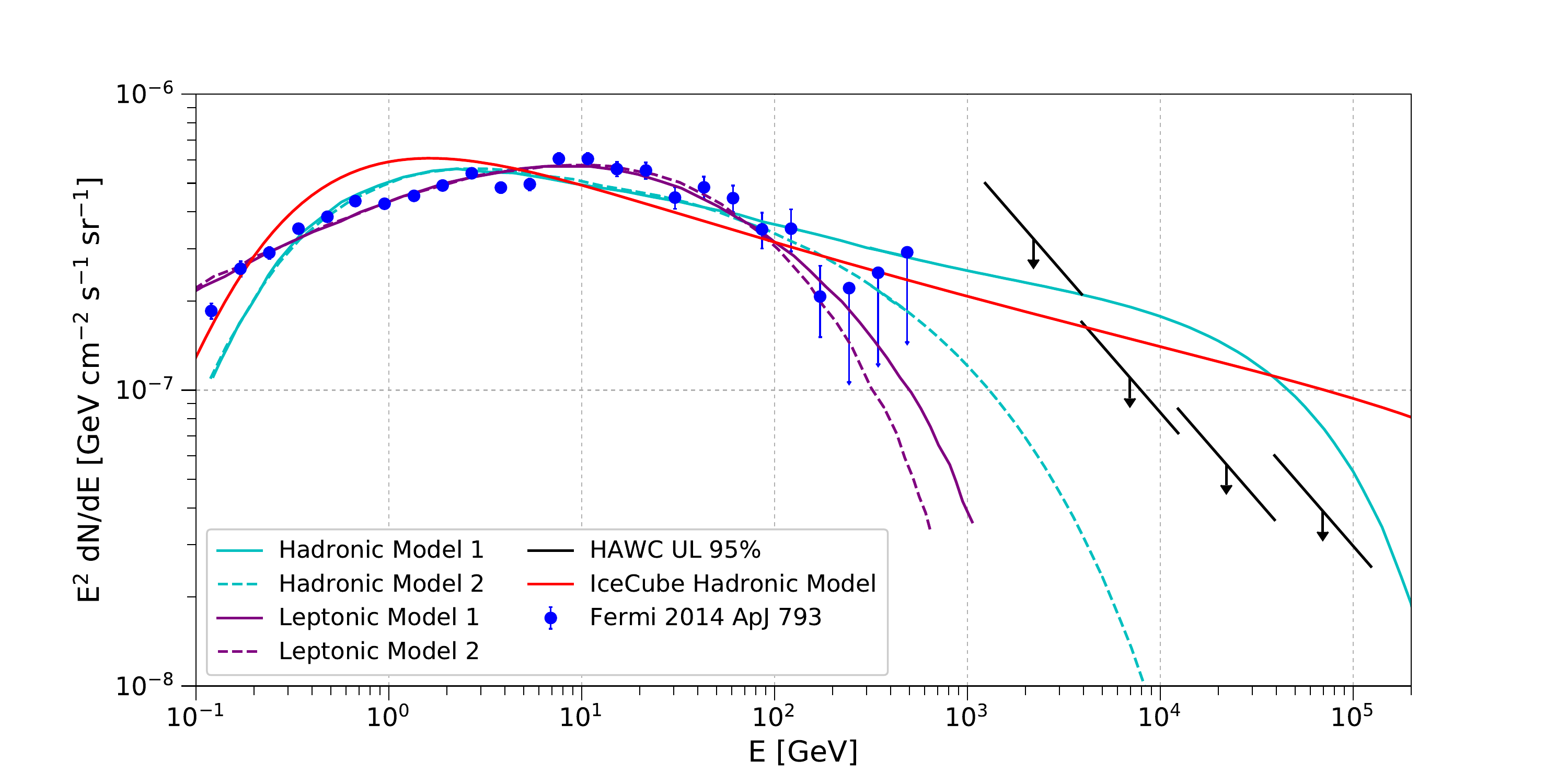}	
	\caption{HAWC upper limits together with the \textit{Fermi} data and gamma-ray production models from ~\cite{Ackermann2014} and ~\cite{Lunardini2}. See table \ref{table:models} for spectral assumptions of these models.}
	\label{fig:ul}
\end{figure*}

Table \ref{table:ulvalues} shows the values of the upper limits and sensitivities for each energy bin. The upper limits obtained from data are consistent with the detection power of HAWC. 
Figure \ref{fig:ul} shows the upper limits together with the flux measurement of the \textit{Fermi} Bubbles made by the \textit{Fermi} Collaboration \citep{Ackermann2014}. 
Different leptonic and hadronic models are also present in Figure \ref{fig:ul}. 

The two leptonic models are obtained from~\cite{Ackermann2014}. In these models, the emission is due to inverse Compton scattering. Two radiation fields are used: the  IRF at 5 kpc above the Galactic Plane and photons from the CMB. The electron spectrum interacting with the radiation fields is modeled as a power-law  with an exponential cutoff.  The spectral index has a value of $2.17\pm0.5^{+0.33}_{-0.89}$ and the cutoff energy is $1.25\pm0.13^{+1.73}_{-0.68}$\,TeV.

The two cyan hadronic models, also obtained from~\cite{Ackermann2014}, assume a power-law and a power-law with cutoff for the injection spectrum of the hadrons.  These protons interact with the ISM producing neutral pions that decay into gamma rays. The spectrum was obtained using the library cparamlib\footnote{\url{https://github.com/niklask/cparamlib}}, which implements the cross sections from \cite{kamae2006}, for the production of gamma rays through hadronic interactions. The spectral index for the power-law is 2.2; the spectral index for the power-law with cutoff is $2.13\pm0.01^{+0.15}_{-0.52}$ with a cutoff energy of $14\pm7^{+6}_{-13}$\,TeV. Using the fit results obtained in ~\cite{Ackermann2014}, we extrapolate the results for the hadronic models above 100\,TeV. The upper limits derived from HAWC data exclude the hadronic injection without a cutoff, that best fits the GeV gamma-ray data, above 3.9\,TeV. 

The hadronic model represented by the red line is obtained from ~\cite{Lunardini2}. This model is the counterpart of a neutrino flux model that best fits the IceCube data. The IceCube data corresponds to five events that are spatially correlated with the \textit{Fermi} Bubbles. The differential flux model was obtained by taking into account the flux from both bubbles. 
Above 10\,TeV, the HAWC upper limits exclude the parent proton spectrum predicted from IceCube data. 

Table \ref{table:models} gives a summary of the different models.

\begin{deluxetable}{ccc}
\tablewidth{0pc}
\tablecaption{Charactersitics of the non-detection: upper limits on the differential flux in four different energy bins. Since the highest energy bin is treated as an overflow bin, only the lower boundary of that energy bin is quoted in order to be conservative.\label{table:ulvalues}}
\tablehead{
\colhead{ Energy Range } & \colhead{ Upper Limits } & \colhead{ Sensitivity } \\
\colhead{ [$\text{TeV}$] } & \colhead{[$ \text{GeV}  \text{cm}^{-2} \text{s}^{-1} \text{sr}^{-1}$]} & \colhead{[$ \text{GeV}  \text{cm}^{-2} \text{s}^{-1} \text{sr}^{-1}$]} }
\startdata
1.2 - 3.9  & 3.0$\times 10^{-7} $ & 3.3$\times 10^{-7} $\\ 
3.9 - 12.4  & 1.0$\times 10^{-7} $ &  1.1$\times 10^{-7} $\\ 
12.4 - 39.1  & 0.5$\times 10^{-7} $  & 0.5$\times 10^{-7} $\\ 
$>$39.1  & 0.4$\times 10^{-7} $  & 0.3$\times 10^{-7} $ \\
\enddata
\end{deluxetable}

\begin{deluxetable*}{ccc}
\tablewidth{0pc}
\tablecaption{Differential flux models for the \textit{Fermi} Bubbles.\label{table:models}}
\tablehead{
\colhead{ Model } & \colhead{ Description } 
}
\startdata
Hadronic Model 1  & $N_p \propto p^{-2.2} $\\ 
Hadronic Model 2  & $N_p \propto p^{-2.1} \exp(-pc/14\,\text{TeV})$\\ 
Leptonic Model 1  & $N_e \propto p^{-2.17} \exp(-pc/1.25\,\text{TeV})$ and IRF at 5kpc\\ 
Leptonic Model 2  & $N_e \propto p^{-2.17} \exp(-pc/1.25\,\text{TeV})$ and CMB \\
IceCube Hadronic Model  & $N_p \propto p^{-2.25} \exp(-pc/30\,\text{PeV})$ \\
\enddata

\end{deluxetable*}

Early reports such as ~\cite{Crocker2011,fujita2013}, presented the possibility of observing TeV gamma rays. The intensity was predicted to be $\leq E^2\,F(\text{TeV})\,\sim\,10^{-6}\,\, \text{GeV}  \text{cm}^{-2} \text{s}^{-1} \text{sr}^{-1}$. The result presented here sets a stricter upper limit. 

The result is not  constraining the main contribution to the spectrum of the \textit{Fermi} Bubbles. Nevertheless our result may imply, for a hadronic model, that there is a cutoff in the proton spectrum. ~\cite{Ackermann2014} showed that GeV gamma-ray spectrum cuts off around 100\,GeV. The cutoff for the parent proton spectrum in this case could be around 1\,TeV ~\citep{Cheng2015}.

As mentioned in Section \ref{subsec:excess}, \cite{fujita2013,Yang2014,Mou2015} propose that the size of the bubbles increases with energy. While defining the search region to be the same as the excess detected at GeV energies is a more conservative approach, it may be interesting to increase the size of the latter in a follow-up analysis.

Increasing the sensitivity at energies $<$1\,TeV is another objective for future analysis. Compared to recent\citep{CrabPaper, CatalogPaper} or future \citep{gemingapaper}  publications of the analysis of HAWC data, this analysis uses only the seven highest event-size bins. At energies $\lesssim$1\,TeV, the large-scale anisotropy signal (or any significant, spatially-extended feature) causes signal contamination in the estimation of the background because the structure takes up a large portion of the field-of-view of HAWC, significantly altering the all-sky rate.  An iterative procedure for the DI method will be followed as explained in  \cite{fiorino2016} and has been shown to remove this artifact.

\section{Conclusions}
\label{sec:Conclusion}
A search of high-energy gamma rays in the Northern \textit{Fermi} Bubble region has been presented by using 290 days of data from the HAWC observatory. No significant excess is found above 1.2\,TeV in the search area and the 95\% C.L. flux upper limits are calculated and compared to the differential sensitivity with $\alpha=0.05$ and $\beta=0.5$.  The upper limits are between 3$\times 10^{-7}$\,$\text{GeV}\, \text{cm}^{-2}\, \text{s}^{-1}\, \text{sr}^{-1}$ and 4$\times 10^{-8}$\,$\text{GeV}\, \text{cm}^{-2}\, \text{s}^{-1} \,\text{sr}^{-1}$ . 
The upper limits, for gamma-ray energies between 3.9\,TeV and 120\,TeV, disfavor the emission of hadronic models that try to explain the GeV gamma-ray emission detected by the \textit{Fermi} LAT. This makes a continuation of the proton injection above 100\,TeV highly unlikely (solid cyan line in Figure \ref{fig:ul}). The HAWC upper limits also disfavor a hadronic injection spectrum derived from IceCube measurements.
The present result does not allow unequivocal conclusions about the hadronic or leptonic origin of the Fermi bubbles though.
A future analysis of HAWC data will include a better sensitivity, especially at lower energies and possibly larger search regions according to the predictions of some theoretical models.

\acknowledgements
We	acknowledge	the	support	from:	the	US	National	Science	Foundation	(NSF);	the	
US	Department	of	Energy	Office	of	High-Energy	Physics;	the	Laboratory	Directed	
Research	and	Development	(LDRD)	program	of	Los	Alamos	National	Laboratory;	
Consejo	Nacional	de	Ciencia	y	Tecnología	(CONACyT),	México	(grants	271051,	
232656,	260378,	179588,	239762,	254964,	271737,	258865,	243290,	132197),	
Laboratorio	Nacional	HAWC	de	rayos	gamma;	L'OREAL	Fellowship	for	Women	in	
Science	2014;	Red	HAWC,	México;	DGAPA-UNAM	(grants	RG100414,	IN111315,	
IN111716-3,	IA102715,	109916,	IA102917);	VIEP-BUAP;	PIFI	2012,	2013,	
PROFOCIE	2014,	2015; the	University	of	Wisconsin	Alumni	Research	Foundation;	
the	Institute	of	Geophysics,	Planetary	Physics,	and	Signatures	at	Los	Alamos	
National	Laboratory;	Polish	Science	Centre	grant	DEC-2014/13/B/ST9/945;	
Coordinación	de	la	Investigación	Científica	de	la	Universidad	Michoacana. Thanks	to	
Luciano	Díaz	and	Eduardo	Murrieta	for	technical	support.
We thank Cecilia Lunardini for providing us with the gamma-ray flux model from IceCube data.


\appendix

\section{Table of variables}\label{app:variables}
\begin{deluxetable}{c|c}[!h]
\tablewidth{0pc}
\tablecaption{Description of the variables used in the analysis}
\tablehead{
\colhead{ Variable  } & \colhead{ Description } 
}
\startdata
$N_i$ & Number of events in sky-map pixel $i$ before gamma-hadron cuts\\ 
$N'_i$ & Number of events in sky-map pixel $i$ after gamma-hadron cuts\\
$\langle N_i \rangle$ & Number of estimated background events in sky-map pixel $i$ before gamma-hadron cuts\\ 
$\langle N'_i \rangle$ & Number of estimated background events in sky-map pixel $i$ after gamma-hadron cuts\\
$E_i$ & Excess above background in sky-map pixel $i$ before gamma-hadron cuts\\
$E'_i$ & Excess above background in sky-map pixel $i$ after gamma-hadron cuts\\
\enddata

\end{deluxetable}

\section{Uncertainty Calculation of G'}\label{app:uncertainty}
The number of gamma rays, as presented in Section \ref{subsec:excess}, is given by
\begin{equation}\label{eq:excess2}
G'_i = \varepsilon_{G,i} G_i,
\end{equation}
where $G_i$ is given by
\begin{equation}
G_i=\frac{E'_i -\varepsilon_{C,i} E_i}{\varepsilon_{G,i}-\varepsilon_{C,i}}.
\end{equation}
The value of $\varepsilon_{C,i}$ is obtained by the equation
\begin{equation}
\varepsilon_{C,i} = \sum_j \langle N'_j \rangle / \sum_j \langle N_j \rangle,
\end{equation}
where $j$ are the pixels in the same HEALPix ring as pixel $i$.

The uncertainty $\delta G_i$ is calculated as
\begin{equation}
\left(\frac{\delta G}{|G|}\right)^2 = \frac{1}{(E'_i - \varepsilon_{C,i}E_i)^2} \left[ \delta E_i ^{'2} + (\varepsilon_{C,i} E_i)^2[(\frac{\delta \varepsilon_{C,i}}{\varepsilon_{C,i}})^2 + (\frac{\delta E_i}{E_i})^2 ]  \right] + \frac{\delta \varepsilon_{C,i}^2}{(\varepsilon_{G,i}-\varepsilon_{C,i})^2}.
\end{equation}

The uncertainties of the different terms are  $\delta E_i ^{'} = \sqrt{\langle N'_i \rangle}$; $\delta E_i  = \sqrt{\langle N_i \rangle}$; and $\delta \varepsilon_{C,i} = | \varepsilon_{C,i}| \sqrt{\alpha_i \left( \frac{1}{\sum_j \langle N'_j \rangle} + \frac{1}{\sum_j \langle N_j \rangle} \right)}$, where $\alpha_i$ is the relative exposure of the observed sky map to the direct integration background. 
It is calculated as $\alpha_i = \Delta \Omega / (\Delta \theta \, \Delta t \, 15^{\degree} \, hr^{-1} \, \cos \delta$), where $\Delta \Omega$ is the pixel area, $\Delta \theta$ is the pixel size, $\Delta t$ is the integration time and $\delta$ is the declination.

The systematic uncertainty on $G = \sum_i G_i$ due to the gamma-ray content in the variable  $\varepsilon_{C,i}$ is estimated.  First we calculate the relative error of the measured $\varepsilon_{C,i}$ to the true value $ \varepsilon^t_{C,i}$ where the superscript is for ``true".
\begin{equation}
\frac{\delta \varepsilon_{C,i}}{\varepsilon^t_{C,i}}=\frac{|\varepsilon_{C,i} - \varepsilon^t_{C,i}|}{\varepsilon^t_{C,i}} = \frac{\sum_j ( \varepsilon_{G,i}-\varepsilon^t_{C,i}) G^I_j}{\varepsilon^t_{C,i} \sum_j ( C^I_j + G^I_j ) } \approx \frac{\sum_j ( \varepsilon_{G,i}) G^I_j}{\varepsilon^t_{C,i} \sum_j ( C^I_j ) } < \frac{\sum_j ( \varepsilon_{G,i}) G_j}{\varepsilon^t_{C,i} \sum_j ( C^I_j ) },
\end{equation}
where the numerator is close to the gamma-ray signal after gamma-hadron cuts and the denominator is close to the isotropic background after gamma-hadron cuts. 
By using the information from \cite{CrabPaper}, the relative error is estimated. 

The total trigger rate for the HAWC observatory is 24kHz for the 2sr field of view.
Assuming that most of the Crab events come from a $1^{\degree}$ radius, we can obtain an estimation of the background rate events from the Crab. 
This background rate is defined as $BR = 24000 (2 \pi ( 1- \cos(1^{\degree}))/2sr$. This background rate is proportional to the isotropic component after gamma-hadron cuts $\varepsilon_C C_I$
The total number of observed events from the Crab is $166.85 \text{ events/transit}$ or $0.00772 \text{Hz}$ for 6 hour/transit. This is proportional to the excess gamma rays after gamma-hadron cuts $\varepsilon_G G^T$. We calculate the following ratio
\begin{equation}
\frac{N' - \langle N' \rangle}{\langle N' \rangle} = \frac{\varepsilon_G G + \varepsilon_C C^I}{\varepsilon_C C^I}. 
\end{equation}
The ratio $C/C_I$ is $O(10^{-4})$\citep{Abeysekara2014}, so the ratio $\frac{N' - \langle N' \rangle}{\langle N' \rangle}$ can be approximated as 
\begin{equation}
\frac{N' - \langle N'\rangle}{\langle N' \rangle} = \frac{\varepsilon_G G }{\varepsilon_C C_I} = 0.6\times10^{-3}. 
\end{equation}

The systematic error in $G_i$ can then be written as
\begin{equation}
(\delta G_i)_{sys} = \frac{\partial G_i}{\partial \varepsilon_{C,i}} \delta \varepsilon_{C,i}
\end{equation}
Assuming gaussian regime, $(\delta G)_{sys} = \sum _i (\delta G_i)^2_{sys}$, where $i$ is pixel number.
The ratio $(\delta G)_{sys.}/(\delta G)_{stat.}$ is of order $O(10^{-4})$, so the systematic uncertainty is ignored.

\bibliography{hfb} 

\begin{thebibliography}{24}
\expandafter\ifx\csname natexlab\endcsname\relax\def\natexlab#1{#1}\fi

\bibitem[{Abeysekara {et~al.}(2013)Abeysekara, Alfaro, Alvarez, Álvarez,
  Arceo, Arteaga-Velázquez, Solares, Barber, Baughman, Bautista-Elivar,
  Belmont, BenZvi, Berley, Rosales, Braun, Caballero-Lopez, Carramiñana,
  Castillo, Cotti, Cotzomi, de~la Fuente, León, DeYoung, Hernandez,
  Diaz-Velez, Dingus, DuVernois, Ellsworth, Fernandez, Fiorino, Fraija,
  Galindo, Garcia-Luna, Garcia-Torales, Garfias, González, González, Goodman,
  Grabski, Gussert, Hampel-Arias, Hui, Hüntemeyer, Imran, Iriarte, Karn,
  Kieda, Kunde, Lara, Lauer, Lee, Lennarz, Vargas, Linares, Linnemann, Longo,
  Luna-García, Marinelli, Martinez, Martínez-Castro, Matthews,
  Miranda-Romagnoli, Moreno, Mostafá, Nava, Nellen, Newbold, Noriega-Papaqui,
  Oceguera-Becerra, Patricelli, Pelayo, Pérez-Pérez, Pretz, Rivière, Ryan,
  Rosa-González, Salazar, Salesa, Sandoval, Santos, Schneider, Silich, Sinnis,
  Smith, Sparks, Springer, Taboada, Toale, Tollefson, Torres, Ukwatta,
  Villaseñor, Weisgarber, Westerhoff, Wisher, Wood, Yodh, Younk, Zaborov,
  Zepeda, \& Zhou}]{Abeysekara2013}
Abeysekara, A., {et~al.} 2013, Astroparticle Physics, 50–52, 26

\bibitem[{Abeysekara {et~al.}(2014)}]{Abeysekara2014}
---. 2014, ApJ, 796, 108

\bibitem[{{Abeysekara} {et~al.}(2017{\natexlab{a}}){Abeysekara}, {Albert},
  {Alfaro}, {et~al.}}]{CatalogPaper}
{Abeysekara}, A.~U., {Albert}, A., {Alfaro}, R., {et~al.} 2017{\natexlab{a}},
  ArXiv e-prints

\bibitem[{{Abeysekara} {et~al.}(2017{\natexlab{b}}){Abeysekara}, {Albert},
  {Alfaro}, {et~al.}}]{CrabPaper}
---. 2017{\natexlab{b}}, ArXiv e-prints

\bibitem[{Ackermann {et~al.}(2014)Ackermann, Albert, {et~al.}}]{Ackermann2014}
Ackermann, M., Albert, A., {et~al.} 2014, ApJ, 793, 64

\bibitem[{Ahlers {et~al.}(2016)Ahlers, BenZvi, Desiati, {et~al.}}]{fiorino2016}
Ahlers, M., BenZvi, S.~Y., Desiati, P., {et~al.} 2016, The Astrophysical
  Journal, 823, 10

\bibitem[{Atkins {et~al.}(2003)Atkins, Benbow, Berley, {et~al.}}]{Atkins2003}
Atkins, R., Benbow, W., Berley, D., {et~al.} 2003, ApJ, 595, 803

\bibitem[{Cheng {et~al.}(2015)Cheng, Chernyshov, Dogiel, \& Ko}]{Cheng2015}
Cheng, K.~S., Chernyshov, D.~O., Dogiel, V.~A., \& Ko, C.~M. 2015, The
  Astrophysical Journal, 804

\bibitem[{Cheng {et~al.}(2011)Cheng, Chernyshov, Dogiel, Ko, \& Ip}]{Cheng2011}
Cheng, K.-S., Chernyshov, D.~O., Dogiel, V.~A., Ko, C.~M., \& Ip, W. 2011, The
  Astrophysical Journal, 731, L17

\bibitem[{Crocker \& Aharonian(2011)}]{Crocker2011}
Crocker, R.~M., \& Aharonian, F. 2011, Physical Review Letters, 106, 101102

\bibitem[{Dobler {et~al.}(2010)Dobler, Finkbeiner, Cholis, Slatyer, \&
  Weiner}]{Dobler2010}
Dobler, G., Finkbeiner, D.~P., Cholis, I., Slatyer, T.~R., \& Weiner, N. 2010,
  ApJ, 717, 825

\bibitem[{Fujita {et~al.}(2013)Fujita, Ohira, \& Yamazaki}]{fujita2013}
Fujita, Y., Ohira, Y., \& Yamazaki, R. 2013, The Astrophysical Journal Letters,
  775, 20

\bibitem[{Fujita {et~al.}(2014)Fujita, Ohira, \& Yamazaki}]{fujita2014}
---. 2014, The Astrophysical Journal, 789, 67

\bibitem[{Gorski {et~al.}(2005)Gorski, Hivon, Banday, {et~al.}}]{Gorski2005}
Gorski, K., Hivon, E., Banday, A., {et~al.} 2005, ApJ, 622, 759

\bibitem[{Guo \& Mathews(2012)}]{Guo2012a}
Guo, F., \& Mathews, W.~G. 2012, The Astrophysical Journal, 756, 181

\bibitem[{Guo {et~al.}(2012)Guo, Mathews, Dobler, \& Oh}]{Guo2012b}
Guo, F., Mathews, W.~G., Dobler, G., \& Oh, S.~P. 2012, The Astrophysical
  Journal, 756, 182

\bibitem[{{HAWC Collaboration}(2017, in preparation)}]{gemingapaper}
{HAWC Collaboration}. 2017, in preparation

\bibitem[{Helene(1983)}]{Helene1983}
Helene, O. 1983, Nucl. Instrum. Methods Phys. Res., 212, 319

\bibitem[{{Kamae} {et~al.}(2006){Kamae}, {Karlsson}, {Mizuno},
  {et~al.}}]{kamae2006}
{Kamae}, T., {Karlsson}, N., {Mizuno}, T., {et~al.} 2006, \apj, 647, 692

\bibitem[{Kashyap {et~al.}(2010)Kashyap, van Dyk, Connors, Freeman,
  Siemiginowska, Xu, \& Zezas}]{Vinay2010}
Kashyap, V.~L., van Dyk, D.~A., Connors, A., Freeman, P.~E., Siemiginowska, A.,
  Xu, J., \& Zezas, A. 2010, The Astrophysical Journal, 719, 900

\bibitem[{Lunardini {et~al.}(2015)Lunardini, Razzaque, \& Yang}]{Lunardini2}
Lunardini, C., Razzaque, S., \& Yang, L. 2015, Phys. Rev. D, 92, 021301

\bibitem[{Mou {et~al.}(2015)Mou, Yuan, Gan, \& Sun}]{Mou2015}
Mou, G., Yuan, F., Gan, Z., \& Sun, M. 2015, The Astrophysical Journal, 811, 37

\bibitem[{Su {et~al.}(2010)Su, Slatyer, \& Finkbeiner}]{Su2010}
Su, M., Slatyer, T.~R., \& Finkbeiner, D.~P. 2010, ApJ, 724, 1044

\bibitem[{Yang {et~al.}(2014)Yang, Aharonian, \& Crocker}]{Yang2014}
Yang, R.-Z., Aharonian, F., \& Crocker, R. 2014, A\&A, 567, 8

\end{thebibliography}

\end{document}